\documentclass[preprintnumbers, nofootinbib,superscriptaddress,10pt]{revtex4}
\usepackage[dvipdfmx]{graphicx} 
\usepackage{amsmath,amssymb,amsfonts,bm} 

  \def\g{\gamma}  \def\d{\delta}     \def\th{\theta}   \def\l{\lambda}  \def\m{\mu} \def\n{\nu}     \def\r{\rho}   \def\t{\tau}       

\def\dg{\dagger}  \def\nn{\nonumber}

\newcommand{\lsp}{ \left ( } \newcommand{\rsp}{ \right ) }
\newcommand{\To}{\Rightarrow}

\renewcommand{\Im}{{\rm Im}\,}

\newcommand{\Diag}[3]{ \begin{pmatrix} #1 & 0 & 0 \\ 0 & #2 & 0 \\ 0 & 0 & #3 \\\end{pmatrix}}

\usepackage{color}

\begin{document}

\title{\large Rephasing invariant structure of CP phase in Fritzsch--Xing parametrization
 \\ for simplified mixing matrices }

\preprint{STUPP-25-292}


\author{Masaki J. S. Yang}
\email{mjsyang@mail.saitama-u.ac.jp}
\affiliation{Department of Physics, Saitama University, 
Shimo-okubo, Sakura-ku, Saitama, 338-8570, Japan}
\affiliation{Department of Physics, Graduate School of Engineering Science,
Yokohama National University, Yokohama, 240-8501, Japan}



\begin{abstract} 

In this paper, we construct an explicit rephasing transformation that converts an arbitrary unitary mixing matrix into the Fritzsch--Xing (FX) parametrization, which is obtained by trivializing arguments of the matrix elements in the third row and third column. 
We further analyze rephasing invariant structure of the FX phase 
$\delta_{\rm FX}$ under an approximation $U_{13}^{e} = 0$, 
where the 1-3 element of  the diagonalization matrix of charged leptons $U^{e}$ is neglected. 
With an additional approximation $U_{23}^{e} = 0$, 
the FX phase becomes highly simplified, reducing to a sum of the neutrino-intrinsic FX phase 
$\delta^{\nu}_{\rm FX}$ and the contribution from the relative phase $\rho'_{1}- \rho'_{2}$ between the lighter 1-2 generations. 
The phase $\delta_{\rm FX}$ for finite $U_{23}^{e}$ is understood as a generalization of the compact expression.
This result covers almost all perturbative calculations of CP phases for the CKM and MNS matrices with hierarchical charged fermions.

\end{abstract} 

\maketitle

\section{Introduction}

The origin and structure of CP phases in the fermion mixing matrices constitute one of the central problems in particle physics, and the Dirac CP phase $\d$ in the lepton mixing matrix provides a direct clue to the underlying new physics.
Since the CP phase depends on parametrizations and a choice of basis, various rephasing invariants, starting from the  Jarlskog invariant \cite{Jarlskog:1985ht}, have been discussed 
\cite{Wu:1985ea, Bernabeu:1986fc, Gronau:1986xb, Branco:1987mj,  Bjorken:1987tr, Nieves:1987pp,  Botella:1994cs, Kuo:2005pf, Jenkins:2007ip, Chiu:2015ega, Denton:2020igp}. 
These studies have mainly evaluated the magnitude of CP violation through quantities proportional to $\sin \d$ in the PDG parametrization \cite{Chau:1984fp}, and did not explicitly address the phase structure itself.

On the other hand, the CP phase can admit another interpretation under different parametrizations, 
potentially revealing more fundamental structures. 
For example, in the original Kobayashi--Maskawa (KM) parametrization of the CKM matrix \cite{Kobayashi:1973fv}, 
the CP phase is close to the maximal value $\pi/2$, and this phenomenological indication has been extensively studied 
 \cite{Koide:2004gj, Koide:2008yu,   Frampton:2010ii,Dueck:2010fa,Frampton:2010uq,Li:2010ae,Qin:2011bq,Zhou:2011xm,Qin:2011ub,Qin:2010hn,Zhang:2012ys,Li:2012zxa,Zhang:2012bk}. 
As another parametrization, the Fritzsch--Xing (FX) parametrization proposed in Ref.~\cite{Fritzsch:1997fw} 
 decomposes the CKM matrix into the lighter 1-2 and the heavier 2-3 generation mixing to describe the characteristic structure of the quark mixing. 
In this parametrization as well, it has been discussed that the FX phase $\d_{\rm FX}$ is also nearly maximal $\pi/2$
\cite{Shin:1985cg, Shin:1985vi, Lehmann:1995br, Kang:1997uv, Mondragon:1998gy, Fritzsch:1999im, Fritzsch:2002ga, Xing:2003yj, Xing:2009eg, Barranco:2010we, Xing:2015sva, Yang:2020qsa, Yang:2020goc, Fritzsch:2021ipb, Awasthi:2022nbi}. 
However, transformations between different parametrizations have only been mentioned implicitly, and the explicit  matrices of transformations have been largely unknown. Moreover, CP phases in various parametrizations have conventionally been obtained through rather cumbersome calculations of matrix elements.

Recently, through studies of rephasing invariants  involving the determinant \cite{Yang:2025hex,Yang:2025cya,Yang:2025law,Yang:2025ftl,Yang:2025dhm,Yang:2025vrs}, 
CP phases and rephasing transformations between various parametrizations are systematically described in terms of  elements of the mixing matrices \cite{Yang:2025dkm, Yang:2025qlg, Yang:2026ulu}.
Within this new framework, the structure of CP phases in the mixing matrix and the transformations among different parametrizations are understood transparently. 

In this paper, we provide an explicit rephasing transformation that converts an arbitrary unitary matrix into the FX parametrization.   
Since this rephasing transformation can be applied not only to the mixing matrix but also to the diagonalization matrices of individual fermions $U^{\n ,e}$, we can  analyze deeper structures of  the CP phase by fermion-specific invariants.  
%

\section{Explicit rephasing transformation of the Fritzsch--Xing parametrization}

The parametrization proposed by Fritzsch and Xing  \cite{Fritzsch:1997fw} decomposes the quark mixings into the lighter 1-2 and the heavier 2-3 mixing, 
based on the characteristic features of the quark mixing. 
\begin{align}
U' & = 
\begin{pmatrix}
c_{u} & s_{u} & 0 \\
- s_{u} & c_{u} & 0 \\
0 & 0 & 1 \\
\end{pmatrix}
\begin{pmatrix}
e^{- i \d_{\rm FX}} & 0 & 0 \\
0 & c_{q} & s_{q} \\
0 & - s_{q} & c_{q}
\end{pmatrix}
\begin{pmatrix}
c_{d} & - s_{d} & 0 \\
s_{d} & c_{d} & 0 \\
0 & 0 & 1 
\end{pmatrix} \nn \\ 
& = 
\begin{pmatrix}
 s_u s_d c_q + c_u c_d e^{-i \delta _{\text{FX}}} & s_u c_d c_q-c_u s_d e^{-i \delta _{\text{FX}}} & s_u s_q \\
 c_u s_d c_q - s_u c_d e^{-i \delta _{\text{FX}}} & c_u c_d c_q + s_u s_d e^{-i \delta _{\text{FX}}}& c_u s_q \\
 -s_d s_q & -c_d s_q & c_q \\
\end{pmatrix} . 
\label{originalFX}
\end{align}
This FX parameterization is often more suitable than the standard PDG representation, because the 1-3 mixing of each fermion can be neglected for hierarchical mass matrices which possess approximate chiral symmetries, as will be justified in the next section. 

As in the PDG parametrization, by rephasing five fermion fields and using the sign freedom of $\d_{\rm FX}$, 
one can choose all $c_f$ and $s_f$ to be positive, i.e., $c_f, s_f > 0$.
The phase structure of this parametrization is specified by the following six conditions 
\begin{align}
\arg U_{e3}' = \arg U_{\m3}' = \arg  U_{\t3}' = \arg [- U_{\t2}' ] = \arg [ - U_{\t1}'] = 0 \, , ~~ 
\arg  \det U' = - \delta_{\rm FX} \, .
\end{align}
We define the rephasing transformation from $U'$ to the mixing matrix $U$ in an arbitrary basis as 
\begin{align}
\begin{pmatrix}
U_{e1} & U_{e2} & U_{e3}  \\
U_{\m1} & U_{\m 2} & U_{\m 3}  \\
U_{\t 1} & U_{\t 2} & U_{\t 3}  \\
\end{pmatrix}
= 
\Diag{e^{i \g_{L1}}}{e^{i \g_{L2}}}{e^{i \g_{L3}}}
\begin{pmatrix}
U_{e1}' & U_{e2}' & |U_{e3}| \\
U_{\m1}' & U_{\m 2}' & |U_{\m 3}| \\
- |U_{\t 1}| & - |U_{\t 2}| & |U_{\t 3}| \\
\end{pmatrix}
\Diag{e^{ - i \g_{R1}}}{e^{ - i \g_{R2}}}{e^{ - i \g_{R3}}} .
\end{align}
The FX phase $\d_{\rm FX}$ is obtained from the determinant and the matrix elements with trivial arguments, 
\begin{align}
\arg \left[ {U_{e3} U_{\m3} U_{\t 1} U_{\t 2} } \right]
 & =  \g_{L1}  + \g_{L2} + 2 \g_{L3} - \g_{R1}  - \g_{R2} - 2 \g_{R3}  \, , \nn \\
 \arg \left[ {U_{e3} U_{\m 3} U_{\t 1} U_{\t 2} \over  \det U } \right] & = 
 \g_{L3}  -  \g_{R3} + \d_{\rm FX}
 ~~ \To ~~ 
\d_{\rm FX} =  \arg \left[ {U_{e3} U_{\m 3} U_{\t 1} U_{\t 2} \over U_{\t 3} \det U } \right] \, . 
\label{FX}
\end{align}
Including the transformation property of the determinant, this is manifestly rephasing invariant and therefore corresponds to an observable. 

The advantage of this representation is that the phase $\d_{\rm FX}^{q}$ is very close to the maximal value $\pi /2$ in the CKM matrix. 
The  parameters quoted from the latest UTfit are \cite{UTfit:2022hsi} 
\begin{align}
\sin \th_{12}^{q} &= 0.22519 \pm 0.00083 \, ,  ~~~ \sin \th_{23}^{q} = 0.04200 \pm 0.00047 \, ,  \nn \\
\sin \th_{13}^{q} &= 0.003714 \pm 0.000092 \, ,  ~~~ \d^{q} = 1.137 \pm 0.022  = 65.15 \pm1.3^{\circ}  \, . 
\end{align}
The numerical evaluation of $\d_{\rm FX}$ at the best-fit point is 
\begin{align}
\d_{\rm FX}^{q} =  \arg \left[ {V_{ub} V_{c b} V_{t d} V_{t s} \over V_{t b} \det V_{\rm CKM} } \right] 
=  93.45^{\circ} \, . 
\label{observe}
\end{align}
Exploring the origin of this large CP phase is theoretically intriguing, 
and many papers have discussed characteristic CP structures and various forms of CP symmetry 
\cite{Shin:1985cg, Shin:1985vi, Lehmann:1995br, Kang:1997uv, Mondragon:1998gy, Fritzsch:1999im, Fritzsch:2002ga, Xing:2003yj, Xing:2009eg, Barranco:2010we, Xing:2015sva, Yang:2020qsa, Yang:2020goc, Fritzsch:2021ipb, Awasthi:2022nbi}.

As in previous studies \cite{Yang:2025dkm, Yang:2025qlg, Yang:2026ulu}, we perform an explicit rephasing transformation by solving for the unknown phases $\g_{Li}$ and $\g_{Ri}$.  
The five conditions to be solved are
\begin{align}
\g_{L1} - \g_{R3} & = \arg U_{e3} \, , ~ 
\g_{L2} - \g_{R3} = \arg U_{\m3} \, , ~ 
\g_{L3} - \g_{R3} = \arg U_{\t3} \, , \nn \\
\g_{L3} - \g_{R2} & = \arg [- U_{\t 2}] \, , ~ 
\g_{L3} - \g_{R1} = \arg [- U_{\t 1}] \, . 
\end{align}
Due to the freedom of the  overall phase, one phase remains unsolved.  
By leaving $\g_{R3}$ as the unsolved variable, $\g_{R3}$ cancels between the left and right phase matrices.  
As a result, the explicit rephasing transformation is given by 
\begin{align}
U =  
\Diag{e^{i \arg U_{e3}  }}{e^{i  \arg U_{\m 3} }}{e^{  i \arg  U_{\t 3} }}
U'
\Diag{e^{i \arg [- U_{\t 1}/ U_{\t 3} ] } }{e^{ i \arg [- U_{\t 2}/ U_{\t 3}] } }{ 1} . 
\label{explicitFX}
\end{align}
We find that the transformation ultimately amounts to trivializing five phases.  
By actually performing the explicit rephasing  and converting to the FX parametrization, 
\begin{align}
 \Diag{e^{- i \arg U_{e3}  }}{e^{- i  \arg U_{\m 3} }}{e^{ - i \arg  U_{\t 3} }}
U
\Diag{e^{ i \arg [- U_{\t 3}/ U_{\t 1} ] } }{e^{ i \arg [- U_{\t 3}/ U_{\t 2}] } }{ 1} 
 = 
\begin{pmatrix}
U_{e1}' & U_{e2}' & |U_{e3}| \\
U_{\m1}' & U_{\m 2}' & |U_{\m 3}|  \\
- |U_{\t 1}| & - |U_{\t 2}| & |U_{\t 3}| \\
\end{pmatrix} , 
\end{align}
the four nontrivial arguments are written by the angles of unitarity triangles 
\begin{align}
\arg U_{e 1}' & = \arg \left[ - {  U_{e 1} U_{\t 3}  \over U_{e3} U_{\t 1}  } \right]  , ~~ 
\arg U_{e 2}' = \arg \left[ - { U_{e 2} U_{\t 3}  \over U_{e3} U_{\t 2}  } \right] , \nn \\
\arg U_{\m 1}' & = \arg \left[ - { U_{\m 1} U_{\t 3} \over U_{\m 3} U_{\t 1} } \right] , ~~ 
\arg U_{\m 2}'  = \arg \left[ - { U_{\m 2} U_{\t 3}  \over U_{\m 3} U_{\t 2}  } \right] . 
\label{nontrivial}
\end{align}
Since $U_{e3}, U_{\m 3}, U_{\t 3}, - U_{\t 2}$ and $- U_{\t 1}$  carry no arguments  in the FX parametrization, 
the validity of relations is manifest from the rephasing invariance. 

Similarly, 
the explicit rephasing transformation to the PDG parametrization $U^{\rm PDG}$ is \cite{Yang:2025dkm}
\begin{align}
U = 
\Diag{e^{i \arg U_{e1} } }{e^{ i  \arg \left[ {\det U \over  U_{e2}  U_{\t 3} } \right]}}{e^{ i \arg \left[ {  \det U \over U_{e2} U_{\m3} } \right] }}
U^{\rm PDG}
\Diag{1}{e^{ i \arg [{U_{e2}\over U_{e1} }] } }{e^{  i \arg \left[ { U_{e2} U_{\m3} U_{\t 3} \over \det U } \right] }} . 
\label{PDG}
\end{align}
The rephasing between the PDG and FX parametrizations is achieved by the product of two diagonal phase matrices. 
The transformation between the two CP phases $\d_{\rm FX}$ and $\d_{\rm PDG}$ are also obtained 
by evaluating them in each other's parametrizations, 
\begin{align}
\d_{\rm FX} & = \arg \left[ {U_{e3} U_{\m 3} U_{\t 1} U_{\t 2} \over U_{\t 3} \det U } \right]
=  - \d_{\rm PDG} + \arg \left[ - { s_{12} s_{23} - c_{12} c_{23} s_{13} \, e^{+ i \d_{\rm PDG}}  \over  c_{12} s_{23} + s_{12} c_{23} s_{13} \,  e^{- i \d_{\rm PDG}}  }\right] , \nn \\
\d_{\rm PDG} & = \arg \left[ {U_{e1} U_{e2} U_{\m3} U_{\t3} \over U_{e3} \det U } \right]
= \d_{\rm FX} + 
\arg \left[ {c_q s_d s_u+c_d c_u e^{-i \delta _{\text{FX}}} \over c_d c_q s_u-c_u s_d e^{+ i \delta _{\text{FX}}} } \right]  . 
\end{align}
Here, $s_{ij}$ and $c_{ij}$ denote the mixing angles in the PDG parametrization.

\section{Rephasing invariant structure of FX phase for simplified mixing matrices}

In this section, we investigate the generic behavior of the FX phase by  rephasing invariants. 
The lepton mixing matrix  $U \equiv U^{e\dagger} U^{\nu}$, defined as the mismatch between the diagonalization matrices $U^{\n , e}$ , is considered under the following approximation.

\begin{description}
\item[\bf Approximation:] 
The 1-3 element of $U^{e}$ is neglected, $U_{13}^{e} = 0$.  

\item[\bf Justification:] 

When the mass matrix $m_{e}$ possesses chiral symmetries for the first and second generations, $m_{e} = D_{L} m_{e} D_{R}$, all mixings and lighter singular values vanish. Here, $D_{L,R} \equiv {\rm diag}(e^{i\phi_{L,R}^{1}}, e^{i\phi_{L,R}^{2}}, 1)$ with phases $\phi_{L,R}^{1,2}$. Although these chiral symmetries are only approximate in reality, the mixing angles are suppressed by powers of the ratios of singular values $m_{ei}/m_{ej}$.

\end{description}

\subsection{The case of $U_{13}^{\n} = U_{13}^{e} = 0$}

Before considering the general situation, we analyze a simpler case with an additional condition $U_{13}^{\nu} = 0$. 
By partially using the inversion formula to make the unitarity constraints explicit, 
\begin{align}
U &= 
\begin{pmatrix}
U_{11}^{e *} & - {U_{12}^e U_{33}^e \over \det U^{e}} & { U_{12}^{e} U_{23}^{e} \over \det U^{e}}  \\[2pt]
U_{12}^{e *} & {U_{11}^e U_{33}^e \over \det U^{e} } & - {U_{11}^{e} U_{23}^{e} \over \det U^{e } } \\[2pt]
0 & U_{23}^{e *} & U_{33}^{e *} \\
\end{pmatrix} 
\begin{pmatrix}
U_{11}^{\n} & U_{12}^{\n} & 0\\[2pt]
- {U_{12}^{\n*} U_{33}^{\n*} \over \det U^{\n *} }  & {U_{11}^{\n*} U_{33}^{\n*} \over \det U^{\n*}} & U_{23}^{\n} \\[2pt]
{ U_{12}^{\n*} U_{23}^{\n*} \over \det U^{\n *}} & - {U_{11}^{\n*} U_{23}^{\n*} \over \det U^{\n * } } & U_{33}^{\n} \\
\end{pmatrix}  \nn \\ 
& = 
\begin{pmatrix}
*  & * & - {U_{12}^e \over \det U^{e} }  (U^{e}_{33} U^{\n}_{23} - U^{e}_{23} U^{\n}_{33} )  \\[2pt]
*  &  * & {U_{11}^e \over \det U^{e} }  (U^{e}_{33} U^{\n}_{23} - U^{e}_{23} U^{\n}_{33} )  \\[2pt]
{U_{12}^{\n*}\over \det U^{\n *} } (U^{e *}_{33} U^{\n *}_{23} -  U^{e *}_{23} U^{\n *}_{33} ) & 
- {U_{11}^{\n*}\over \det U^{\n *} } (U^{e *}_{33} U^{\n *}_{23} -  U^{e *}_{23} U^{\n *}_{33} )& U^{e *}_{23} U^{\n}_{23} +  U^{e *}_{33} U^{\n}_{33}
\end{pmatrix} . 
\end{align}
Here, the matrix elements denoted by $*$ are not relevant for the phase calculation.
Since some of factors are complex conjugates, the FX phase $\d_{\rm FX}$ becomes remarkably simplified as 
\begin{align}
\d_{\rm FX} & = \arg \bigg[ \frac{ U^{e}_{11} U^{\n *}_{11}  U^{e}_{12} U^{\n *}_{12} 
\left( U^{e *}_{33} U^{\n *}_{23} - U^{e *}_{23} U^{\n *}_{33} \right)^2 
(U^{e}_{33} U^{\n}_{23} - U^{e}_{23} U^{\n}_{33} )^2  }
{\det U^{e} \det U^{\n*} ( U^{e *}_{23}U^{\n}_{23} + U^{e *}_{33}U^{\n}_{33}  )} \bigg ] \nn \\
& = \arg \bigg[ \frac{U^{e}_{11} U^{e}_{12} U^{\n *}_{11} U^{\n *}_{12} }
{\det U^{e} \det U^{\n*} ( U^{e *}_{23} U^{\n}_{23} + U^{e *}_{33} U^{\n}_{33} )} \bigg] \, . 
\end{align}
On the other hand, for $U_{13}^{\n,e} = 0$,
 all CP phases of $U^{\n,e}$ can be removed by the PDG rephasing transformation~(\ref{PDG}). 
Thus, the physical relative phases between them are 
\begin{align}
\r_{1} -\r_{2}  & =  \arg \left[  { U^{e *}_{11}  U^{e *}_{12}  U^{e *}_{33} \over \det U^{e *}  } { U^{\n}_{11} U^{\n}_{12}  U^{\n}_{33} \over \det U^{\n} } \right] ,  ~~
 \r_{2} - \r_{3} =  \arg \left[ { U^{e *}_{23} U^{\n}_{23}   \over U^{e*}_{33} U^{\n}_{33}  } \right]  .  
 \label{relative}
\end{align}
Separating the phases explicitly, 
\begin{align}
\d_{\rm FX}  =  \arg \left[  { U^{e}_{11}  U^{e}_{12}  U^{e}_{33} \over \det U^{e}  } { U^{\n *}_{11} U^{\n *}_{12}  U^{\n *}_{33} \over \det U^{\n *} } \right]
- 
 \arg \left[ 1 + {U^{e *}_{23} U^{\n}_{23} \over U^{e *}_{33}  U^{\n}_{33}}  \right]   \, . 
 \label{semigeneral}
\end{align}
By setting the standard parametrization,
\begin{align}
U = 
\begin{pmatrix}
c_{12}^{e} & -s_{12}^{e} & 0 \\
s_{12}^{e} & c_{12}^{e} & 0 \\
0 & 0 & 1 \\
\end{pmatrix}
\begin{pmatrix}
1 & 0 & 0 \\
0 & c_{23}^{e} & - s_{23}^{e} \\
0 & s_{23}^{e} & c_{23}^{e}
\end{pmatrix}
\Diag{e^{i \r_{1}}}{e^{ i \r_{2}}}{e^{ i \r_{3}}} 
\begin{pmatrix}
1 & 0 & 0 \\
0 & c_{23}^{\n} & s_{23}^{\n} \\
0 & - s_{23}^{\n} & c_{23}^{\n}
\end{pmatrix}
\begin{pmatrix}
c_{12}^{\n} & s_{12}^{\n} & 0 \\
- s_{12}^{\n} & c_{12}^{\n} & 0 \\
0 & 0 & 1 
\end{pmatrix} , 
\label{para1}
\end{align}
it indeed agrees with the result  of invariants 
\begin{align}
\d_{\rm FX} = \r_{2} - \r_{1} 
- \arg \left[ c^e_{23} c^{\n}_{23}  + e^{i (\rho _{2} - \rho_{3} )} s^e_{23} s^{\n}_{23}   \right ] \, . 
\end{align}

In particular, in the limit where $|U_{23}^{e}| = s_{23}^{e}$ is sufficiently small, $\d_{\rm FX}$ is further simplified 
by the expansion $\arg [1 + x] = \Im \log [1 + x] \simeq \Im x$, 
\begin{align}
\d_{\rm FX} & \simeq \arg \left[  { U^{e}_{11}  U^{e}_{12}  U^{e}_{33} \over \det U^{e}  } { U^{\n *}_{11} U^{\n *}_{12}  U^{\n *}_{33} \over \det U^{\n *} } \right]  - \Im \left[  {U^{e *}_{23} U^{\n}_{23} \over U^{e *}_{33}  U^{\n}_{33}}  \right]  \nn \\
& =  \r_{2} - \r_{1} -  \sin (\rho _{2} - \rho_{3} ) {s^e_{23} s^{\n}_{23} \over c^e_{23} c^{\n}_{23} } \, . 
\end{align}
Since Eq.~(\ref{para1}) with $s^{e}_{23} = 0$ reduces to the original parametrization~(\ref{originalFX}), 
direct field redefinitions yield the same result $- \d_{\rm FX} = \r_{1} - \r_{2}$. 

\subsection{The case of $U_{23}^{e} = U_{13}^{e} = 0$}

Neglecting $U_{13}^{\nu}$ may not be a good approximation, because the actual lepton mixing matrix has $|U_{e3}| \simeq 0.15$. 
Therefore, it is meaningful to evaluate the CP phase for finite $U_{13}^{\nu}$.
Before proceeding to the general situation, let us consider another simplified scenario, 
in which $U_{23}^{e} = 0$ is imposed by sacrificing the condition $U_{13}^{\nu} = 0$. The mixing matrix is then given by
\begin{align}
U &= 
\begin{pmatrix}
U_{11}^{e *} & - {U_{12}^e U_{33}^e \over \det U^{e}} & 0 \\[2pt]
U_{12}^{e *} & {U_{11}^e U_{33}^e \over \det U^{e} } & 0 \\[2pt]
0 & 0 & U_{33}^{e *} \\
\end{pmatrix} 
\begin{pmatrix}
U_{11}^{\n} & U_{12}^{\n} & U_{13}^{\n} \\[2pt]
U_{21}^{\n} & U_{22}^{\n}  & U_{23}^{\n} \\[2pt]
U_{31}^{\n} & U_{32}^{\n} & U_{33}^{\n} \\
\end{pmatrix} 
 = 
\begin{pmatrix}
*  & * & U_{11}^{e *} U_{13}^{\n} - {U_{12}^e U_{33}^e \over \det U^{e}} U_{23}^{\n} \\[2pt]
*  &  * &  U_{12}^{e *} U_{13}^{\n} + {U_{11}^e U_{33}^e \over \det U^{e} }U_{23}^{\n}   \\[2pt]
U_{33}^{e *} U_{31}^{\n} & U_{33}^{e *} U_{32}^{\n} &U^{e *}_{33} U^{\n}_{33}
\end{pmatrix} , 
\end{align}
and the phase $\d_{\rm FX}$ is found to be
\begin{align}
\d_{\rm FX} & = \arg 
\left[ {
(U_{11}^{e *} U_{13}^{\n} - {U_{12}^e U_{33}^e \over \det U^{e}} U_{23}^{\n})
\over 
(U_{12}^{e *} U_{13}^{\n} + {U_{11}^e U_{33}^e \over \det U^{e} }U_{23}^{\n})^{*}}
{ U_{33}^{e *} U_{31}^{\n}  U_{33}^{e *} U_{32}^{\n} \over U^{e *}_{33} U^{\n}_{33} \det U^{e *} \det U^{\n}} \right] . 
\label{d24}
\end{align}
It seems to be better to separate the FX phase of neutrinos
\begin{align}
\d_{\rm FX} & 
= 
\arg
\left[ 
{(U_{11}^{e *}  - {U_{12}^e U_{33}^e \over \det U^{e}} {U_{23}^{\n} \over U_{13}^{\n}} ) \over 
(U_{12}^{e *} {U_{13}^{\n} \over U_{23}^{\n}} + {U_{11}^e U_{33}^e \over \det U^{e} } )^{*} }
{ U_{33}^{e *}  \over  \det U^{e *} } \right] + \d^{\n}_{\rm FX} ,  ~~~ 
\d^{\n}_{\rm FX} \equiv \arg \left[ { U_{13}^{\n} U_{23}^{\n} U_{31}^{\n} U_{32}^{\n} \over U^{\n}_{33} \det   U^{\n}} \right] . 
\label{semigeneral2}
\end{align}

In the explicit rephasing to the FX representation~(\ref{explicitFX}), 
the relative phase between the 1-2 generations of  neutrinos is
$\r_{1}^{\n} - \r_{2}^{\n} = \arg [{ U^{\n}_{13} / U^{\n}_{23} } ]$.
Thus, the relative phase between the two fermions is 
\begin{align}
\r_{1}' - \r_{2}' =  \arg \left[  { U^{e *}_{11}  U^{e *}_{12}  U^{e *}_{33} \over \det U^{e *}  } { U^{\n}_{13} \over U^{\n}_{23} } \right] 
\equiv \arg R_{12}' \, . 
\end{align}
Even in this case, $\d_{\rm FX}$ is separated by contributions of the two phases $\d^{\n}_{\rm FX}$ and $\r'_{1} - \r'_{2}$; 
\begin{align}
\d_{\rm FX} -  \d^{\n}_{\rm FX}  =  \arg
\left[ 
{({U_{11}^{e *} U_{12}^{e *} U_{33}^{e *}  \over  \det U^{e *}} {U_{13}^{\n} \over U_{23}^{\n}}  - { |U_{12}^e U_{33}^e|^{2} }  )
\over ({U_{11}^e U_{12}^e U_{33}^e \over \det U^{e} } {U_{13}^{\n *} \over U_{23}^{\n *}}+ |U_{12}^{e }|^{2} {|U_{13}^{\n}|^{2} \over |U_{23}^{\n}|^{2} } )^{*}}
  \right]
 = \arg \left[ { R_{12}' - { |U_{12}^e U_{33}^e|^{2} }  \over 
 |U_{23}^{\n}|^{2} R_{12}' + |U_{12}^{e } U_{13}^{\n}|^{2} }   \right] . 
\end{align}

To express this result in terms of the relative phase,
we parameterize the mixing matrix as,
\begin{align}
U & = 
\begin{pmatrix}
c_{12}^{e} & -s_{12}^{e} & 0 \\
s_{12}^{e} & c_{12}^{e} & 0 \\
0 & 0 & 1 \\
\end{pmatrix}
\Diag{e^{i \r'_{1}}}{e^{ i \r'_{2}}}{e^{ i \r'_{3}}} 
\begin{pmatrix}
c_{1}^{\n} & s_{1}^{\n} & 0 \\
- s_{1}^{\n} & c_{1}^{\n} & 0 \\
0 & 0 & 1 \\
\end{pmatrix}
\begin{pmatrix}
e^{- i \d^{\n}_{\rm FX}} & 0 & 0 \\
0 & c_{2}^{\n} & s_{2}^{\n} \\
0 & - s_{2}^{\n} & c_{2}^{\n}
\end{pmatrix}
\begin{pmatrix}
c_{3}^{\n} & - s_{3}^{\n} & 0 \\
s_{3}^{\n} & c_{3}^{\n} & 0 \\
0 & 0 & 1 
\end{pmatrix} .
\label{para2}
\end{align}
The evaluation  by rephasing invariants agrees with that of parameterization 
\begin{align}
\d_{\rm FX} =  \delta^{\nu}_{\rm FX} + 
\arg \left[
\frac{ -c_1^{\n} s^e_{12} + e^{i (\r'_{1} - \r'_{2})} s_1^{\n} c^e_{12}}
{ s_1^{\n} s^e_{12} +  e^{i (\r'_{1} - \r'_{2})} c_1^{\n} c^e_{12} } \right] \, . 
\end{align}
It is interesting that the CP phase is unaffected by $c^{\n}_{2,3}$ and $s^{\n}_{2,3}$.

\subsection{The general case of $U_{13}^{e} = 0$}

Based on these behaviors, we proceed to a general calculation. 
The mixing matrix is
\begin{align}
U &= 
\begin{pmatrix}
U_{11}^{e *} & - {U_{12}^e U_{33}^e \over \det U^{e}} & { U_{12}^{e} U_{23}^{e} \over \det U^{e}}  \\[2pt]
U_{12}^{e *} & {U_{11}^e U_{33}^e \over \det U^{e} } & - {U_{11}^{e} U_{23}^{e} \over \det U^{e } } \\[2pt]
0 & U_{23}^{e *} & U_{33}^{e *} \\
\end{pmatrix} 
\begin{pmatrix}
U_{11}^{\n} & U_{12}^{\n} & U_{13}^{\n} \\[2pt]
U_{21}^{\n} & U_{22}^{\n} & U_{23}^{\n} \\[2pt]
U_{31}^{\n} & U_{32}^{\n} & U_{33}^{\n} \\
\end{pmatrix} \nn \\
&  = 
\begin{pmatrix}
*  & * & U_{11}^{e *} U_{13}^{\n} - {U_{12}^e U_{33}^e \over \det U^{e}} U_{23}^{\n} + { U_{12}^{e} U_{23}^{e} \over \det U^{e}} U_{33}^{\n}  \\[2pt]
*  &  * &  U_{12}^{e *} U_{13}^{\n} + {U_{11}^e U_{33}^e \over \det U^{e} }U_{23}^{\n}  - {U_{11}^{e} U_{23}^{e} \over \det U^{e } } U_{33}^{\n} \\[2pt]
U_{23}^{e *} U_{21}^{\n} + U_{33}^{e *} U_{31}^{\n} & U_{23}^{e *} U_{22}^{\n}  +  U_{33}^{e *} U_{32}^{\n} &
U_{23}^{e *} U_{23}^{\n} +U^{e *}_{33} U^{\n}_{33}
\end{pmatrix} , 
\end{align}
and the most general $\d_{\rm FX}$ in the present approximation is 
\begin{align}
\d_{\rm FX} = 
\arg \left [ {U_{23}^{e *} U_{21}^{\n} + U_{33}^{e *} U_{31}^{\n} \over
 (U_{23}^{e *} U_{22}^{\n}  +  U_{33}^{e *} U_{32}^{\n} )^{*}  }
 { U_{11}^{e *} U_{13}^{\n} - {U_{12}^e  \over \det U^{e}} (U_{33}^e U_{23}^{\n} - U_{23}^{e}U_{33}^{\n}) \over 
( U_{12}^{e *} U_{13}^{\n} + {U_{11}^e \over \det U^{e} } (U_{33}^e U_{23}^{\n}  -  U_{23}^{e}U_{33}^{\n}) )^{* }}
 { \det U^{\n *} \det U^{e} \over U_{23}^{e *} U_{23}^{\n} +U^{e *}_{33} U^{\n}_{33}}
 \right ] . 
 \label{general}
\end{align}
In the limits $U_{13}^{\n} \to 0$ or $U_{23}^{e} \to 0$, this expression includes the two simpler cases discussed previously. 

Although this expression is a function of the three phases $\d_{\rm FX}^{\n}$, $\r'_{1} - \r'_{2}$, and $\r'_{2} - \r'_{3}$,
 it may not be constructive to show that explicitly.
The point is that, the nontrivial arguments of $U_{21}^{\n}$ and $U_{22}^{\n}$ in the FX parameterization~(\ref{nontrivial}) appear in the calculation for finite $U^{\n}_{13}$.
These arguments are functions of $\d^{\n}_{\rm FX}$ after the rephasing~(\ref{explicitFX}).

If $|U_{23}^{e}|$ is sufficiently small, we can perturbatively expand the expression of  $\d_{\rm FX}$.
By separating terms that does not involve $U_{23}^{e}$,
\begin{align}
\d_{\rm FX} & = 
\arg \left [ { U_{33}^{e *} U_{31}^{\n} \over
 (  U_{33}^{e *} U_{32}^{\n} )^{*}  }
 { U_{11}^{e *} U_{13}^{\n} - {U_{12}^e U_{33}^e \over \det U^{e}}  U_{23}^{\n}  \over 
( U_{12}^{e *} U_{13}^{\n} + {U_{11}^e U_{33}^e \over \det U^{e} }  U_{23}^{\n}  )^{* }}
 { \det U^{\n *} \det U^{e} \over U^{e *}_{33} U^{\n}_{33}}
 \right ] \nn \\
 & + 
 \arg \left [ { 1 + {U_{23}^{e *} U_{21}^{\n} \over U_{33}^{e *} U_{31}^{\n}  }\over
 (1 + {U_{23}^{e *} U_{22}^{\n} \over  U_{33}^{e *} U_{32}^{\n} } )^{*}  }
 { 1+  { U_{12}^{e} U_{23}^{e}  U_{33}^{\n} \over  \det U^{e} U_{11}^{e *} U_{13}^{\n} -  U_{12}^e U_{33}^e  U_{23}^{\n}  } \over 
(  1 - { U_{11}^{e} U_{23}^{e}  U_{33}^{\n} \over \det U^{e} U_{12}^{e *} U_{13}^{\n} + U_{11}^e U_{33}^e U_{23}^{\n} } )^{* }}
 { 1 \over 1+ {U_{23}^{e *} U_{23}^{\n} \over U^{e *}_{33} U^{\n}_{33}} }
 \right ] . 
\end{align}
The first term is equal to Eq.~(\ref{d24}). 
Applying  the approximation $\arg [1 + x]  \simeq \Im x$  for the second term, we obtain
\begin{align}
\d_{\rm FX} &\simeq \d_{\rm FX}^{\n} + \arg \left[ { R_{12}' - { |U_{12}^e U_{33}^e|^{2} }  \over 
 |U_{23}^{\n}|^{2} R_{12}' + |U_{12}^{e } U_{13}^{\n}|^{2} }   \right] 
  + \Im \left [ {U_{23}^{e *} \over U_{33}^{e *} } \lsp { U_{21}^{\n} \over  U_{31}^{\n} } + { U_{22}^{\n} \over  U_{32}^{\n} }  - { U_{23}^{\n} \over U^{\n}_{33}} \rsp \right ] \nn \\
& + \Im \left [  { U_{12}^{e} U_{23}^{e}  U_{33}^{\n} \over  \det U^{e} U_{11}^{e *} U_{13}^{\n} -  U_{12}^e U_{33}^e  U_{23}^{\n}  } 
 - { U_{11}^{e} U_{23}^{e}  U_{33}^{\n} \over \det U^{e} U_{12}^{e *} U_{13}^{\n} + U_{11}^e U_{33}^e U_{23}^{\n} } \right ]  . 
\end{align}
The expression covers behaviors of the Dirac CP in most models with the chiral symmetries, 
and they do not depend on a particular parameterization because of the rephasing invariance. 
As in the previous cases, one can also adopt a suitable parameterization to express the general result. 
However, in the conventional $\sin \d$-type expressions of invariants \cite{Yang:2024ulq, Yang:2025yst},  
such sums of multiple phases are cumbersome, and understanding of the global behavior will be difficult.

The same analysis is also applied to the CKM matrix $V_{\rm CKM} \equiv U_{u}^{\dg} U_{d}$ by the replacement $\nu, e \to d, u $. 
The approximation also holds in many situations, 
because the up-type fermions have more hierarchical mass singular values and 
a relation $|U_{13}^{u}| \ll |U_{13}^{d}|$ is expected from the chiral symmetries.
Therefore, this result provides an almost general behavior of CP violation for hierarchical  mass matrices of charged fermions diagonalized perturbatively.

\subsection{Outlook for DUNE and T2HK: CKM-like  mixing of charged leptons } 

We discuss what physical implications can be drawn from past and future measurements of the CP violation.
First, let us examine the CP phase already observed in quark mixing.
By choosing an appropriate basis, elements of both 
$U_{u,d}$ are taken to be of the same order as those of $V_{\rm CKM}$ without loss of generality.
In other words, neglecting $\mathcal{O}(1)$ coefficients and taking $\lambda = 0.22$, the magnitudes of the matrix elements are approximately as follows:
\begin{align}
U_{u,d} \sim 
\begin{pmatrix}
1 & \l & \l^{3}  \\
\l &  1 & \l^{2} \\
\l^{3} & \l^{2} & 1 \\
\end{pmatrix} .
\end{align}

In the simplified scenario discussed above, $U_{13}^{u} = U_{13}^{d} = 0$, 
the CP phase for quarks $\delta^{q}_{\rm FX}$ reduces to Eq.~(\ref{semigeneral}), 
\begin{align}
\d_{\rm FX}^{q} &= \arg \left[  { U^{u}_{11}  U^{u}_{12}  U^{u}_{33} \over \det U^{u}  } { U^{d *}_{11} U^{d *}_{12}  U^{d *}_{33} \over \det U^{d *} } \right]
- 
 \arg \left[ 1 + {U^{u *}_{23} U^{d}_{23} \over U^{u *}_{33}  U^{d}_{33}}  \right] \nn \\
 & = \arg \left[  { U^{u}_{11}  U^{u}_{12}  U^{u}_{33} \over \det U^{u}  } { U^{d *}_{11} U^{d *}_{12}  U^{d *}_{33} \over \det U^{d *} } \right] + O (\l^{4}) \, . 
\end{align}
Thus, the observed nearly maximal FX phase~(\ref{observe}) implies that the relative phase between the lighter 
 generations $\rho_{1} - \rho_{2}$~(\ref{relative}) is almost maximal 
\cite{Shin:1985cg, Shin:1985vi, Lehmann:1995br, Kang:1997uv, Mondragon:1998gy, Fritzsch:1999im, Fritzsch:2002ga, Xing:2003yj, Xing:2009eg, Barranco:2010we, Xing:2015sva, Yang:2020qsa, Yang:2020goc, Fritzsch:2021ipb, Awasthi:2022nbi}. 

Next, we investigate leptonic CP violation that can be observed in future experiments.
With some grand unified theory in mind, 
$U^{e}$ has a CKM-like structure 
$|U^{e}_{12}| \lesssim 0.2$, $|U^{e}_{23}| \lesssim 0.05$, and $|U^{e}_{13}| \lesssim 0.01$.
The observation $|U_{e3}| \simeq 0.15$ implies that the contribution of $U^{e}_{13}$ is at the level of about 6\%, 
which is comparable to the future experimental uncertainty. 
The approximation is justified, and the perturbative treatment for $U^{e}_{23}$ also retains sufficient accuracy. 

Similarly, under the simplified condition $U_{13}^{e} = U_{13}^{\nu} = 0$, the second term in Eq.~(\ref{semigeneral}) can be neglected for $|U_{23}^{e}| \lesssim 0.05$. 
If the first term is as large as that of the CKM matrix due to some grand unified relations, 
the CP phase is expected to be large $\delta_{\rm FX} \sim \delta^{q}_{\rm FX} \simeq \pi /2$, 
and be observed with about three years of data from DUNE and T2HK \cite{DUNE:2020jqi,Hyper-KamiokandeProto-:2015xww}.
Deviations from the expected value can be understood as a perturbation from a finite $U_{13}^{\nu}$.
However, since the CP phase is independent of the absolute values of the matrix elements, conditions for the three observed mixing angles  are required separately.

\section{Summary}

In this paper, we construct an explicit rephasing transformation that converts an arbitrary unitary mixing matrix into the Fritzsch--Xing (FX) parametrization, which is obtained by trivializing arguments of the matrix elements in the third row and third column. 
We further analyze the rephasing invariant structure of the FX phase 
$\d_{\rm FX}$ under an approximation $U_{13}^{e} = 0$, 
where the 1-3 element of the diagonalization matrix of charged leptons $U^{e}$ is neglected. 
With an additional approximation $U_{23}^{e} = 0$, 
the FX phase becomes highly simplified, reducing to  a sum of the neutrino-intrinsic FX phase 
$\d^{\n}$ and the contribution from the relative phase $\r'_{1}- \r'_{2}$ between the lighter 1-2 generations. 
The FX phase for finite $U_{23}^{e}$ is understood as a generalization of the compact expression.
This result covers almost all perturbative calculations of CP phases for the CKM and MNS matrices with hierarchical charged fermions.
This approach provides a transparent interpretation of the origin of observable CP violation, 
 and can offer new insights into flavor model building and CP violation of grand unified theories.

\section*{Acknowledgment}

The study is partly supported by the MEXT Leading Initiative for Excellent Young Researchers Grant Number JP2023L0013.


\end{document}